\documentclass{iau} 
\usepackage{graphicx}
\usepackage{natbib}
\bibpunct{(}{)}{;}{a}{}{,} % to follow the A&A style

\newcommand{\aap}{A\&A}
\newcommand{\pasp}{PASP}
\newcommand{\apj}{ApJ}
\newcommand{\aj}{AJ}
\newcommand{\apjs}{ApJS}
\newcommand{\nat}{Nature}
\newcommand{\mnras}{MNRAS}
\newcommand{\rmxaa}{Rev. Mexicana AyA}

\title[SF and H$_2$ in XMP galaxies] %% give here short title %%
{Star Formation and Molecular Gas in Extremely Metal-poor Galaxies:\\ Insights from the Thermal Balance in the Neutral Gas}

\author[Vianney Lebouteiller]   %% give here short author list %%
{Vianney Lebouteiller$^1$}
\affiliation{$^1$AIM, CEA, CNRS, Universit\'e Paris-Saclay, Universit\'e Paris Diderot, Sorbonne Paris Cit\'e, F-91191 Gif-sur-Yvette, France \\ email: {\tt vianney.lebouteiller@cea.fr}} %\\[\affilskip]
\pubyear{2018}
\volume{344}  %% insert here IAU Symposium No.
\jname{Dwarf Galaxies: From the Deep Universe to the Present}
\editors{A.C. Editor, B.D. Editor \& C.E. Editor, eds.}
\begin{document}

\maketitle

\begin{abstract}
The apparent lack of cold molecular gas in blue compact dwarf (BCD) galaxies is at variance
with their intense star-formation episode. The CO molecule, often used a tracer of H$_2$ through a conversion
function, is selectively photodissociated in dust-poor environments and, as a result, a potentially large fraction of H$_2$ is expected to reside in the so-called CO-dark gas, where it could be traced instead by infrared cooling lines [CI], [CII], and [OI].
Although the fraction of CO-dark gas to total molecular gas is
in theory expected to be relatively large in metal-poor galaxies, many uncertainties remain due to the difficulty in identifying the main heating mechanism associated to the cooling lines observed in such galaxies. \\
Investigations of the \textit{Herschel} Dwarf Galaxy Survey (DGS; \citealt{Madden2013a}) show that the heating mechanism in the neutral gas of BCDs cannot be dominated by the photoelectric effect on dust grains below some threshold metallicity due to the low abundance of dust and polycyclic
aromatic hydrocarbons, implying that other heating mechanisms need to be invoked, along with a new interpretation of the corresponding infrared line diagnostics. In the study presented here and in \cite{Lebouteiller2017a}, we use optical and infrared lines to constrain the physical conditions in the HII region + HI region of the BCD I\,Zw\,18 ($18$\,Mpc; $\approx2\%$ solar metallicity) within a consistent photoionization and photodissociation model. We show that the HI region is entirely heated by a single ultraluminous X-ray source with important consequences on the applicability of [CII] to trace the star-formation rate and to trace the CO-dark gas. We derive stringent upper limits on the size of H$_2$ clumps that may be detected in the future with JWST and IRAM/NOEMA. We also show that the nature of the X-ray source can be inferred from the corresponding signatures in the ISM. Finally we speculate that star formation may be quenched in extremely metal-poor dwarf galaxies due to X-ray photoionization.
\end{abstract}

\firstsection % if your document starts with a section,
\section{Context}

The apparent lack of H$_2$ in low-metallicity galaxies has led to the hypothesis that H$^0$ may participate in the star-formation process, with H$_2$ forming only at the onset of star-formation once the gas is cold enough (e.g., \citealt{Glover2012a,Krumholz2012a}). It is also possible that the H$_2$ content (inferred from dust and a dust-to-gas mass ratio or from CO) and the star-formation rate (SFR) are not coeval quantities, with the H$_2$ readily destroyed in the aftermath of the starburst or because it is produced and rapidly consumed. Finally, it is possible that little molecular gas results in a large SFR (i.e., large efficiency; \citealt{Turner2015a}) or that a large reservoir of CO-dark molecular gas exists \citep{Grenier2005a,Madden1997a}.

The [CII] line at $158\mu$m (along with [OI]) is an important cooling line in the neutral gas, detected at high-$z$ (e.g., \citealt{Lagache2018a,Aravena2016a}) and thought to be a tracer of this so-far hidden molecular gas reservoir, especially in low-metallicity environments. The interpretation and the origin of the [CII] emission in metal-poor galaxies is an important challenge to understand how the ISM structure evolves with metallicity and what diagnostics can be applied to high-z galaxies.

\section{Models}

The BCD I\,Zw\,18 is one of the most metal-poor star-forming galaxy in the nearby Universe, making it an ideal test object to examine [CII] emission in an extreme environment. This galaxy harbors an ultraluminous X-ray source \citep{Kaaret2013a}, which is a rather common property for low-metallicity star-forming galaxies, as there is not only a well-known $L_{\rm X}$-SFR relationship but this relationship steepens at low metallicity \citep{Linden2010a,Brorby2015a}. The X-ray source in I\,Zw\,18 bears potentially important consequences for the heating of the gas, as suggested in \cite{Pequignot2008a}, who investigated the optical lines and \textit{Spitzer} mid-infrared lines and suggested that the neutral gas should be heated through photoionization by X-rays.

In \cite{Lebouteiller2017a} we use improved \textit{Spitzer} measurements, the recent \textit{Herschel} detections of [CII] and [OI] \citep{Cormier2015a}, and the dust properties derived in \cite{RemyRuyer2014a} in a Cloudy model \citep{Ferland2017a} that reflects the ISM topology used in \cite{Pequignot2008a}. In short, the ISM is represented as radiation-bounded shells embedded in a more diffuse matter-bounded medium in pressure equilibrium. 

\section{Gas heating and molecular gas}

From the Cloudy models, we infer that the neutral gas heating from the photoelectric effect (PE) alone is negligible ($\lesssim10\%$) even if the dust-to-gas mass ratio is scaled with the metallicity (while the integrated dust-to-gas mass ratio is much lower). This is a remarkable result that contrasts with the predominance of the PE in the Milky Way (e.g., \citealt{Weingartner2001a}). It is also different from what is obtained in more moderately metal-poor environments such as the Magellanic Clouds ($\sim20-50$\% solar metallicity) where there seems to be a compensation between the lower PE heating rate (due to the lower PAH abundance) and the larger PE heating efficiency (due to the lower dust grain charging itself due to the larger mean free path of UV photons in a dust-poor medium) \citep{Israel2011a}. In such a metal-poor galaxy as I\,Zw\,18, the possibly larger PE heating efficiency cannot compensate anymore the overall low heating rate. 

Instead, our models suggest that the main heating mechanism in the neutral gas is photoionization by X-rays, producing an X-ray dominated region where the gas is warm (few $100$ K) and partially ionized (up to $\sim0.1-1$\%). The electron fraction is consistent with what we measured using far-UV absorption line from the C$^+$ fine-structure level \citep{Lebouteiller2013a}.

We examine the molecular gas content under two hypotheses, either as uniformly distributed in the radiation-bounded sector or as tiny dense clumps. Based on the $160\mu$m and CO non-detections, the size of such hypothetical clumps is $\lesssim1$\,pc, which is reminiscent of the CO clumps discovered in the WLM galaxy with ALMA ($\approx10\%$ solar metallicity; \citealt{Rubio2015a}). However, I\,Zw\,18 being about $10$ times further away, it will remain difficult if not impossible to detect CO clumps at such distance. We show that such clumps are better identified through their warm H$_2$ layer (as seen with JWST or ground-based $8$-m class telescopes). These hypothetical clumps could potentially harbor a large quantity of molecular gas ($\lesssim10^7$\,M$_\odot$, i.e., almost as much as the mass of H$^0$), such that an extreme star-formation efficiency is actually not required.

Even if such clumps do exist, we show that most ($\gtrsim85\%$) of the [CII] and [OI] emission actually originate from the more diffuse medium within the radiation-bounded sector where they trace an almost purely atomic gas. We show that most of the molecular gas in the clumps is actually CO-bright, but as explained above, detecting these clumps through CO may be near impossible. We show that the X$_{\rm CO}$ factor required to get to the total H$_2$ mass is compatible with the extrapolation of the \cite{Schruba2012a} relationship. 

\section{Nature of the X-ray source}

The intrinsic X-ray spectrum is that of an accretion disk as suggested in \cite{Kaaret2013a}, corresponding to a near maximum stellar-mass black hole for a direct collapse ($\approx80$\,M$_\odot$). We show that it is possible to infer the X-ray spectrum using the ISM signatures as constraints. The main ISM tracers that bear the signature of the X-ray source are the [NeV] lines in the optical and mid-infrared. By parameterizing the X-ray spectrum as a family of blackbodies, we are able to narrow down the possible X-ray spectra compatible with the XMM-\textit{Newton} observations using the absorbing H column density from far-UV observations and using the [NeV] upper limits. We confirm henceforth the accretion-disk model used in \cite{Kaaret2013a} while illustrating how a realistic ISM topology together with a multi-phase modeling can provide enough constraints to infer the input radiation field in particular in the soft X-ray domain.

\section{Implications}

Overall, we argue that there is a compensation between the lower PE heating rate at low-metallicity (due to the lower dust-to-gas mass ratio) and the larger heating from X-ray photoionization (due to the enhanced abundance and luminosity of high-mass X-ray binaries). Since the star-forming gas reservoir is kept warm on long time scales and large spatial scales, we also speculate that star formation in extremely metal-poor objects could be quenched because of X-ray sources. It is important that simulations at zero- or near zero-metallicity include X-rays sources as potential heating sources.

%\cite[Ott (1993)]{Ott93}

%\begin{figure}[b]
%% \vspace*{-2.0 cm}
%\begin{center}
%% \includegraphics[width=3.4in]{Path.eps} 
%% \vspace*{-1.0 cm}
% \caption{Path of pre-solar grains from their stellar sources to the laboratory.}
%   \label{fig1}
%\end{center}
%\end{figure}

\begin{discussion}

  \discuss{Mirabel}{Have you calculated the fraction of X-ray photons escaping the galaxy? It could bear some important cosmological implications regarding the reionization of the Universe by dwarf galaxies.}

  \discuss{Lebouteiller}{We haven't calculated this fraction. The models are currently adapted to the HII region and does not include the extended HI envelope, although it is feasible. As far as the potential implications are concerned, we do not know yet whether IZw18 is representative of high-z dwarf galaxies, in particular regarding X-ray and ISM properties. The dust-to-gas mass ratio and the ISM topology are probably more important parameters than metallicity alone. A sample of local dwarfs is under examination. }

\end{discussion}

\end{document}